\documentclass[a4paper,11pt]{article}
\usepackage{pos}
\usepackage{subfigure}
\usepackage{hyperref}
\usepackage{lineno}

\title{Collective phenomena study in small systems using the bulk of particle production in high-multiplicity pp collisions  at $\sqrt{s}=$ 13 TeV with ALICE}

\author*[a,b]{Rajendra Nath Patra}
\author{for the ALICE Collaboration}

\affiliation[a]{INFN Bari, Bari, Italy}

\affiliation[b]{University of Jammu, Jammu, India}

\emailAdd{rajendra.nath.patra@cern.ch, rajendrapatra07@gmail.com}

\abstract{The heavy-ion collisions (A--A) at the Large Hadron Collider (LHC) energies have confirmed the production of the quark-gluon plasma (QGP), a new state of nuclear matter where quarks and gluons are deconfined. The light-flavour hadrons ($\pi$, K, p), constitute the bulk of the produced particles, carry useful information of the collision geometry, collective behaviour and thermal property of the QGP. The measurements of light-flavour hadron production in small collision systems (pp and p--A) at the LHC energies have shown the onset of collective phenomena (e.g. radial flow and long-range correlations) that resemble what is typically observed in nucleus-nucleus collisions and attributed to the formation of a deconfined system of quarks and gluons. 
	 
The new results of the identified light-flavour particle production measured in high-multiplicity triggered pp collisions at $\sqrt{s}=13$~TeV of ALICE Run 2 will be presented in search of collective behaviour in small collision systems. The  transverse momenta $p_{\rm T}$-spectra of the identified particles show hardening at the mid-$p_{\rm T}$. The mean transverse momenta ($\langle p_{\rm T} \rangle$) are shown as a function of charged-particle multiplicity. The ratios of $p_{\rm T}$-spectra and the ratios of the integrated yields of kaon- and proton-to-pion are also presented and compared with published results. 
	}

\FullConference{The European Physical Society Conference on High Energy Physics (EPS-HEP2023)\\
 21-25 August 2023\\
Hamburg, Germany\\}

\begin{document}
\maketitle
%

\newcommand{\pp}           {pp\xspace}
\newcommand{\ppbar}        {\mbox{$\mathrm {p\overline{p}}$}\xspace}
\newcommand{\XeXe}         {\mbox{Xe--Xe}\xspace}
\newcommand{\PbPb}         {\mbox{Pb--Pb}\xspace}
\newcommand{\pA}           {\mbox{pA}\xspace}
\newcommand{\pPb}          {\mbox{p--Pb}\xspace}
\newcommand{\AuAu}         {\mbox{Au--Au}\xspace}
\newcommand{\dAu}          {\mbox{d--Au}\xspace}
\newcommand{\CuCu}         {\mbox{Cu--Cu}\xspace}

\newcommand{\s}            {\ensuremath{\sqrt{s}}\xspace}
\newcommand{\st}            {\ensuremath{\sqrt{s \mathrm{t}}}\xspace}
\newcommand{\snn}          {\ensuremath{\sqrt{s_{\mathrm{NN}}}}\xspace}
\newcommand{\pt}           {\ensuremath{p_{\rm T}}\xspace}
\newcommand{\meanpt}       {$\langle p_{\mathrm{T}}\rangle$\xspace}
\newcommand{\ycms}         {\ensuremath{y_{\rm CMS}}\xspace}
\newcommand{\ylab}         {\ensuremath{y_{\rm lab}}\xspace}
\newcommand{\etarange}[1]  {\mbox{$\left | \eta \right |~<~#1$}}
\newcommand{\yrange}[1]    {\mbox{$\left | y \right |~<$~0.5}}
\newcommand{\dndeta}       {\ensuremath{\mathrm{d}N_\mathrm{ch}/\mathrm{d}\eta}\xspace}
\newcommand{\avdndeta}     {\ensuremath{\langle\dndeta\rangle}\xspace}
\newcommand{\dNdy}           {\ensuremath{\mathrm{d}N_\mathrm{ch}/\mathrm{d}y}\xspace}
\newcommand{\dNdyy}         {\ensuremath{\mathrm{d}N/\mathrm{d}y}\xspace}
\newcommand{\dNdptdy} {\ensuremath{\mathrm{d}N/\mathrm{d}p_{\rm T}\mathrm{d}y}\xspace}
\newcommand{\Npart}        {\ensuremath{N_\mathrm{part}}\xspace}
\newcommand{\Ncoll}        {\ensuremath{N_\mathrm{coll}}\xspace}
\newcommand{\dEdx}         {\ensuremath{\textrm{d}E/\textrm{d}x}\xspace}
\newcommand{\RpPb}         {\ensuremath{R_{\rm pPb}}\xspace}
\newcommand{\RAA}         {\ensuremath{R_{\rm AA}}\xspace}

\newcommand{\nineH}        {$\sqrt{s}~=~0.9$~Te\kern-.1emV\xspace}
\newcommand{\seven}        {$\sqrt{s}~=~7$~Te\kern-.1emV\xspace}
\newcommand{\twoH}         {$\sqrt{s}~=~0.2$~Te\kern-.1emV\xspace}
\newcommand{\twosevensix}  {$\sqrt{s}~=~2.76$~Te\kern-.1emV\xspace}
\newcommand{\five}         {$\sqrt{s}~=~5.02$~Te\kern-.1emV\xspace}
\newcommand{\twosevensixnn}{$\sqrt{s_{\mathrm{NN}}}~=~2.76$~Te\kern-.1emV\xspace}
\newcommand{\fivenn}       {$\sqrt{s_{\mathrm{NN}}}~=~5.02$~Te\kern-.1emV\xspace}
\newcommand{\LT}           {L{\'e}vy-Tsallis\xspace}
\newcommand{\GeVc}         {Ge\kern-.1emV/$c$\xspace}
\newcommand{\MeVc}         {Me\kern-.1emV/$c$\xspace}
\newcommand{\TeV}          {Te\kern-.1emV\xspace}
\newcommand{\GeV}          {Ge\kern-.1emV\xspace}
\newcommand{\MeV}          {Me\kern-.1emV\xspace}
\newcommand{\GeVmass}      {Ge\kern-.2emV/$c^2$\xspace}
\newcommand{\MeVmass}      {Me\kern-.2emV/$c^2$\xspace}
\newcommand{\lumi}         {\ensuremath{\mathcal{L}}\xspace}

\newcommand{\ITS}          {\rm{ITS}\xspace}
\newcommand{\TOF}          {\rm{TOF}\xspace}
\newcommand{\ZDC}          {\rm{ZDC}\xspace}
\newcommand{\ZDCs}         {\rm{ZDCs}\xspace}
\newcommand{\ZNA}          {\rm{ZNA}\xspace}
\newcommand{\ZNC}          {\rm{ZNC}\xspace}
\newcommand{\SPD}          {\rm{SPD}\xspace}
\newcommand{\SDD}          {\rm{SDD}\xspace}
\newcommand{\SSD}          {\rm{SSD}\xspace}
\newcommand{\TPC}          {\rm{TPC}\xspace}
\newcommand{\TRD}          {\rm{TRD}\xspace}
\newcommand{\VZERO}        {\rm{V0}\xspace}
\newcommand{\VZEROA}       {\rm{V0A}\xspace}
\newcommand{\VZEROC}       {\rm{V0C}\xspace}
\newcommand{\Vdecay} 	   {\ensuremath{V^{0}}\xspace}

\newcommand{\ee}           {\ensuremath{e^{+}e^{-}}} 
\newcommand{\pip}          {\ensuremath{\pi^{+}}\xspace}
\newcommand{\pim}          {\ensuremath{\pi^{-}}\xspace}
\newcommand{\kap}          {\ensuremath{\rm{K}^{+}}\xspace}
\newcommand{\kam}          {\ensuremath{\rm{K}^{-}}\xspace}
\newcommand{\pbar}         {\ensuremath{\rm\overline{p}}\xspace}
\newcommand{\kzero}        {\ensuremath{{\rm K}^{0}_{\rm{S}}}\xspace}
\newcommand{\lmb}          {\ensuremath{\Lambda}\xspace}
\newcommand{\almb}         {\ensuremath{\overline{\Lambda}}\xspace}
\newcommand{\Om}           {\ensuremath{\Omega^-}\xspace}
\newcommand{\Mo}           {\ensuremath{\overline{\Omega}^+}\xspace}
\newcommand{\X}            {\ensuremath{\Xi^-}\xspace}
\newcommand{\Ix}           {\ensuremath{\overline{\Xi}^+}\xspace}
\newcommand{\Xis}          {\ensuremath{\Xi^{\pm}}\xspace}
\newcommand{\Oms}          {\ensuremath{\Omega^{\pm}}\xspace}
\newcommand{\degree}       {\ensuremath{^{\rm o}}\xspace}
\newcommand{\kstar}        {\ensuremath{\rm {K}^{\rm{* 0}}}\xspace}
\newcommand{\phim}        {\ensuremath{\phi}\xspace}
\newcommand{\pik}          {\ensuremath{\pi\rm{K}}\xspace}
\newcommand{\kk}          {\ensuremath{\rm{K}\rm{K}}\xspace}
\newcommand{\kskm}{$\mathrm{K^{*0}/K^{-}}$}
\newcommand{\phikm}{$\mathrm{\phi/K^{-}}$}
\newcommand{\phixi}{$\mathrm{\phi/\Xi}$}
\newcommand{\phiom}{$\mathrm{\phi/\Omega}$}
\newcommand{\xiphi}{$\mathrm{\Xi/\phi}$}
\newcommand{\omphi}{$\mathrm{\Omega/\phi}$}
\newcommand{\kstf} {K$^{*}(892)^{0}~$}
\newcommand{\phf} {$\mathrm{\phi(1020)}~$}
\newcommand{\dd}{\ensuremath{\mathrm{d}}}
\newcommand{\mT}{\ensuremath{m_{\mathrm{T}}}\xspace}
\newcommand{\krr}{\ensuremath{\kern-0.09em}}
\section{Introduction}
The ultimate goal of the ALICE experiment at the CERN LHC is to characterize the properties of the quark-gluon plasma (QGP) \cite{Burza_2018}, the state of nuclear matter at extremely high temperature and pressure where quarks and gluons are deconfined \cite{ALICE_perf_2008, ALICE_perf_2014}. The ALICE experiment has studied the properties of the QGP at the LHC energies using heavy-ion collisions (i.e., Pb--Pb, Xe--Xe), while pp collisions are typically used as a reference. The manifestation of the collective phenomena such as radial flow, long-range correlation and anisotropic flow \cite{PbPb_5020GeV_2020, XeXe_5440GeV_2021, PbPb_2760GeV_2013, ATLAS_pp_2016, pp_pPb_PbPb_XeXe_flow_2019} is attributed at most to the phase of collisions where the QGP medium is created. In particular, the measurement of transverse momentum $p_{\rm T}$-spectra of identified and unidentified particles is sensitive to the collective properties of the QGP fireball created in the heavy-ion collisions \cite{PbPb_5020GeV_2020, XeXe_5440GeV_2021}. The ALICE experiment, dedicated to the QGP study, presents excellent tracking performances, combined with excellent particle identification (PID) using different sub-detectors over a wide range of transverse momentum. Using various PID techniques, it is possible to measure identified hadron spectra in pp, p--Pb, Pb--Pb and Xe--Xe collisions over a transverse momentum range of $0.1<p_{\rm T}<20$~GeV/{\it c}.

The recent observations in small systems (pp, p--Pb) of the radial flow, strangeness enhancement, long-range correlations and anisotropic flow with increasing multiplicity \cite{pPb_5020GeV_2014,pp_13TeV_2020} require a comprehensive explanation on the mechanisms behind these phenomena in small systems. Moreover, one can ask if it is possible to have a unified description of the collective phenomena observed in pp, p--Pb and Pb--Pb collisions. The goal of this analysis is to study transverse momentum spectra of the identified hadrons at high-multiplicity (HM) pp collisions at $\sqrt{s}$ = 13 TeV for investigating the collective phenomena in small systems, which might help in answering the previous question. 

\section{Analysis method}
The analysis is performed on the Run 2 data collected for pp collisions at $\sqrt{s}$ = 13 TeV using the ALICE detector at the LHC. The analysis exploits the central barrel detectors and their excellent PID and tracking capability down to a low momenta range. The most central Inner Tracking System (ITS) detector, a six cylindrical layers silicon detector, is used for the primary vertexing and also capable of PID at low transverse momentum. The Time Projection Chamber (TPC) is used for PID using the $\rm{d}E/\rm{d}x$ measurement. The tracking of the charged particles is performed using the combined information of the ITS and TPC detectors. The Time Of Flight (TOF) detector is used for the PID measurement at a higher $p_{\rm T}$ range than that where the TPC is ineffective. In addition, the kinks topology technique is used for the PID of pion and kaon inside the TPC detector. The V0M detector, which complies of V0A and V0C, is placed at both side of the interaction points for triggering and multiplicity estimation. The data used in this analysis are selected using the high-multiplicity trigger obtained from the V0M signals corresponding to the most central 0.1\% of the minimum bias (MB) events. The rapidity region $|y|<0.5$ was chosen for the track selections.

Three high-multiplicity classes are defined as 0-0.01\% (HM-1), 0.01-0.05\% (HM-2) and 0.05-0.1\% (HM-3) of the MB events corresponding to the average charged-particle multiplicity density ($\langle {\rm d} N_{\rm ch}/{\rm d} \eta \rangle$) equals to 35.82, 32.21, 30.13 respectively \cite{pp_13TeV_multiplicity_2021}. 

\section{Results}
\begin{figure}[h]
	\centering
	\vspace{-3 mm}
	\subfigure[$\pi^+ + \pi^-$]
	{
	\includegraphics[width=0.45\textwidth]{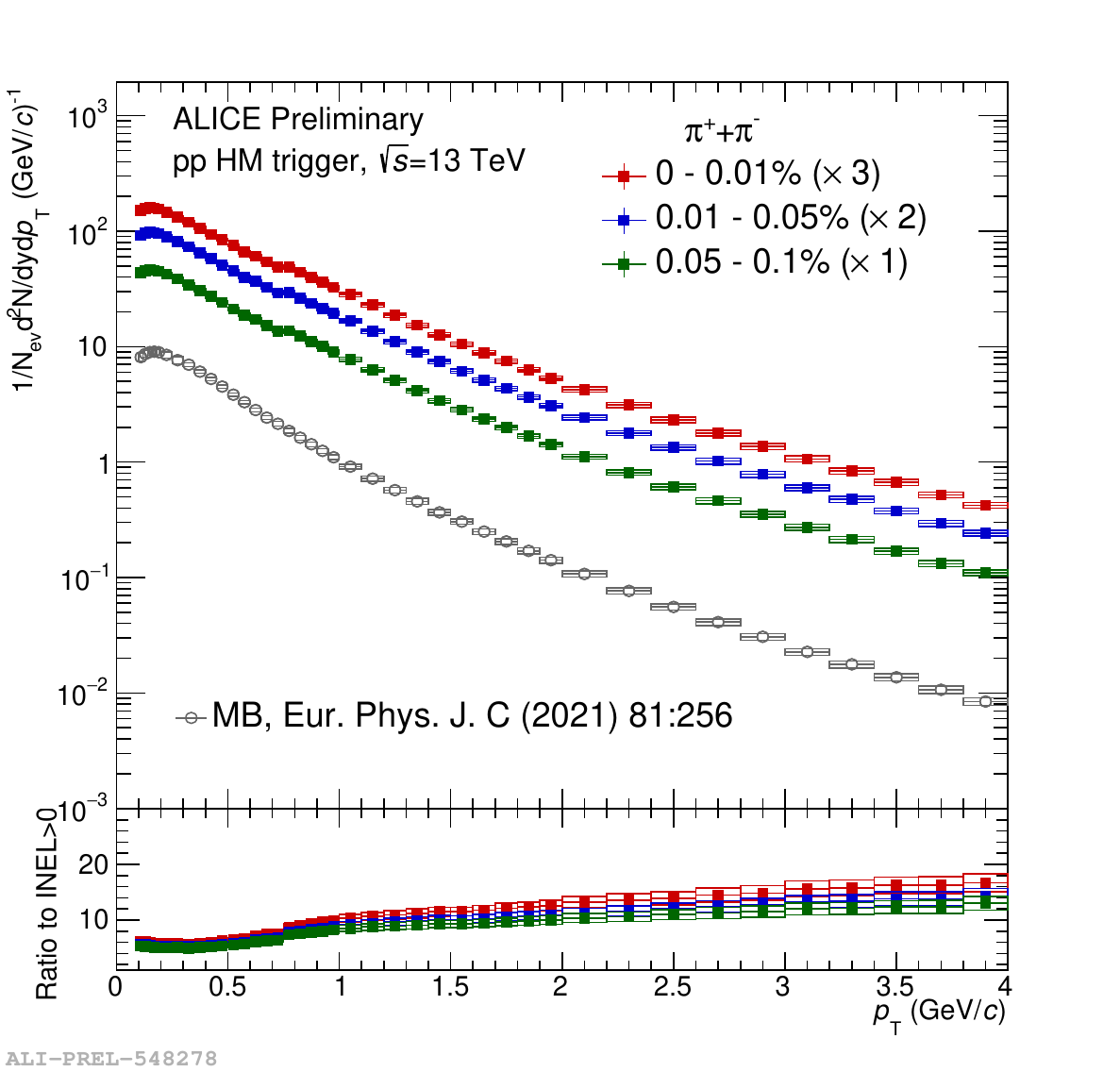}
	}
	\vspace{-3 mm}
	\subfigure[$\rm K^+ + \rm K^-$]
	{
	\includegraphics[width=0.45\textwidth]{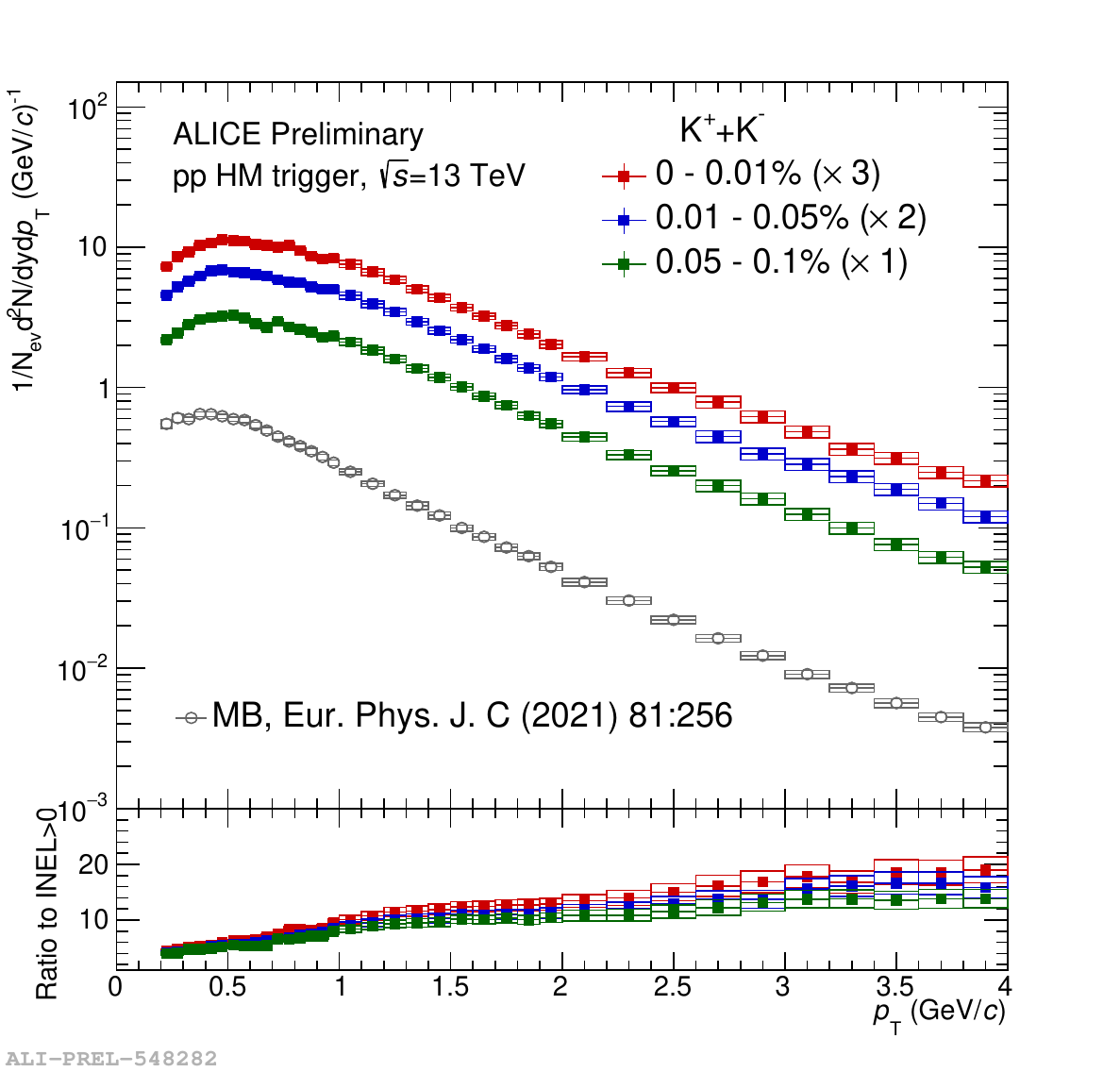}
	}

	\subfigure[$\rm p + \rm \bar{p}$]
	{
	\includegraphics[width=0.45\textwidth]{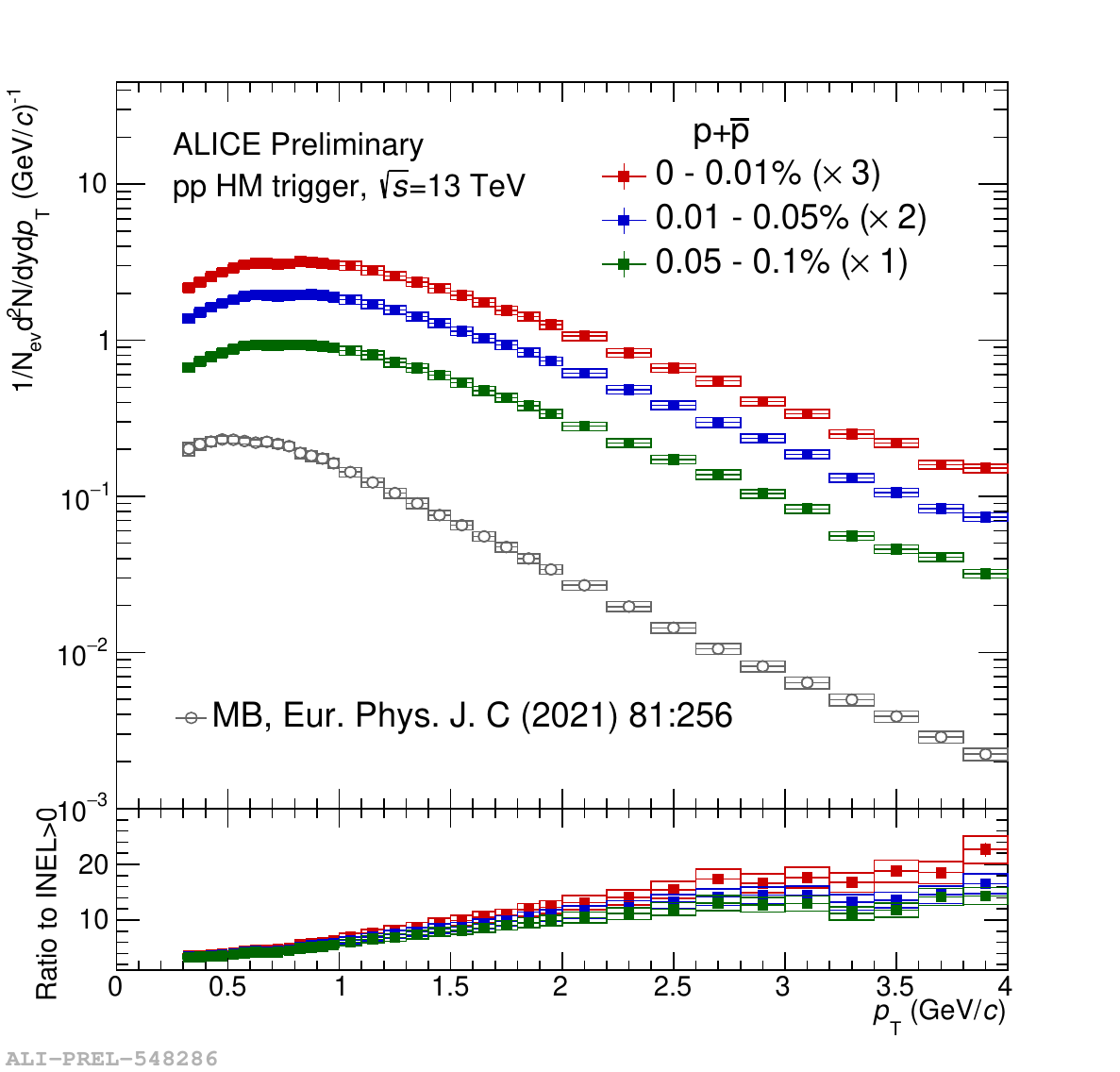}
	}
	\caption{Transverse momentum spectra of $\pi$, K and p for three different high-multiplicity classes and reference minimum bias result \cite{pp_7_13_TeV_MB_2021}. The spectra are scaled to improve the visibility. The ratio to INEL>0 are shown at bottom panels.}
	\label{fig:pi_spectra}
\end{figure}

The $p_{\rm T}$-differential spectra of the $\pi$, K and p of the three HM classes in pp collisions at $\sqrt{s}$ = 13~TeV are shown in Fig.~\ref{fig:pi_spectra}. The reference minimum bias (MB) spectra of INEL>0 are also shown along with the data from ref.~\cite{pp_7_13_TeV_MB_2021}. The bottom panels of the figure show the ratios of the HM spectra to the INEL>0 class. It is found that the spectra become harder with increasing multiplicity and the effect is more pronounced for protons. Such observation, i.e the hardening of $p_{\rm T}$-spectra as multiplicity increases and becomes more prominent for heavier particles, was reported previously in pp, p--Pb and Pb--Pb collisions and it is attributed to the hydrodynamical flow \cite{pp_13TeV_2020, pPb_5020GeV_2014, PbPb_5020GeV_2020}. 

\begin{figure}[h]
	\centering
		\vspace{-3 mm}
	\includegraphics[width=0.9\textwidth]{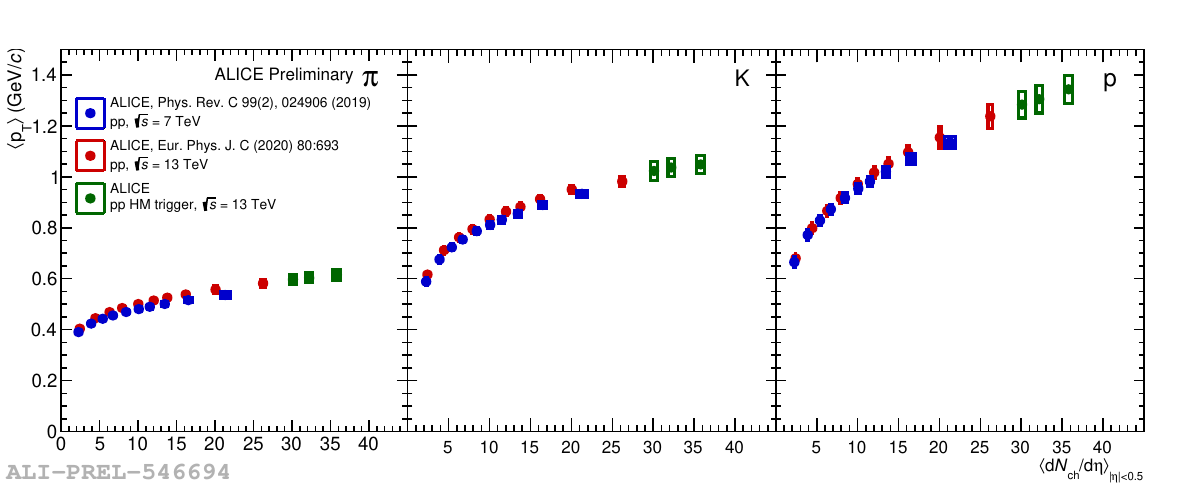}
	
	\caption{Mean transverse momentum $\langle p_{\rm T} \rangle$ as a function of charged-particle multiplicity density ($\langle {\rm d} N_{\rm ch}/{\rm d} \eta \rangle$) of $\pi$, K and p in high-multiplicity pp collisions at $\sqrt{s}$ = 13 TeV and compared to pp results at $\sqrt{s}$ = 7 and 13 TeV \cite{pp_13TeV_2020, pp_7_TeV_2019}.}
	\label{fig:mean_pT}
\end{figure} 

The mean transverse momentum $\langle p_{\rm T} \rangle$ as a function of charged-particle multiplicity density $\langle {\rm d} N_{\rm ch}/{\rm d} \eta \rangle$ is shown in Fig.~\ref{fig:mean_pT} and the results are compared with those obtained for pp collisions at $\sqrt{s}$ = 7 and 13~TeV \cite{pp_13TeV_2020, pp_7_TeV_2019}. A continuously increasing trend of $\langle p_{\rm T} \rangle$ with increasing $\langle {\rm d} N_{\rm ch}/{\rm d} \eta \rangle$ can be observed and the rise is steeper for higher hadron masses. Similar observations
have been reported in pp \cite{CMS_pp_2012} and p–Pb \cite{pPb_5020GeV_2014} collisions previously at lower energies. The effect is consistent with higher radial flow with larger multiplicity and the effect is stronger for particles with larger rest mass. The radial flow-like effect observed on $\langle p_{\rm T} \rangle$ measurement in pp system is qualitatively consistent with the results found in Pb--Pb collisions but at different multiplicity region \cite{PbPb_5020GeV_2020}. 
\begin{figure}[h]
	\centering
	\vspace{-3 mm}
	\includegraphics[width=0.45\textwidth]{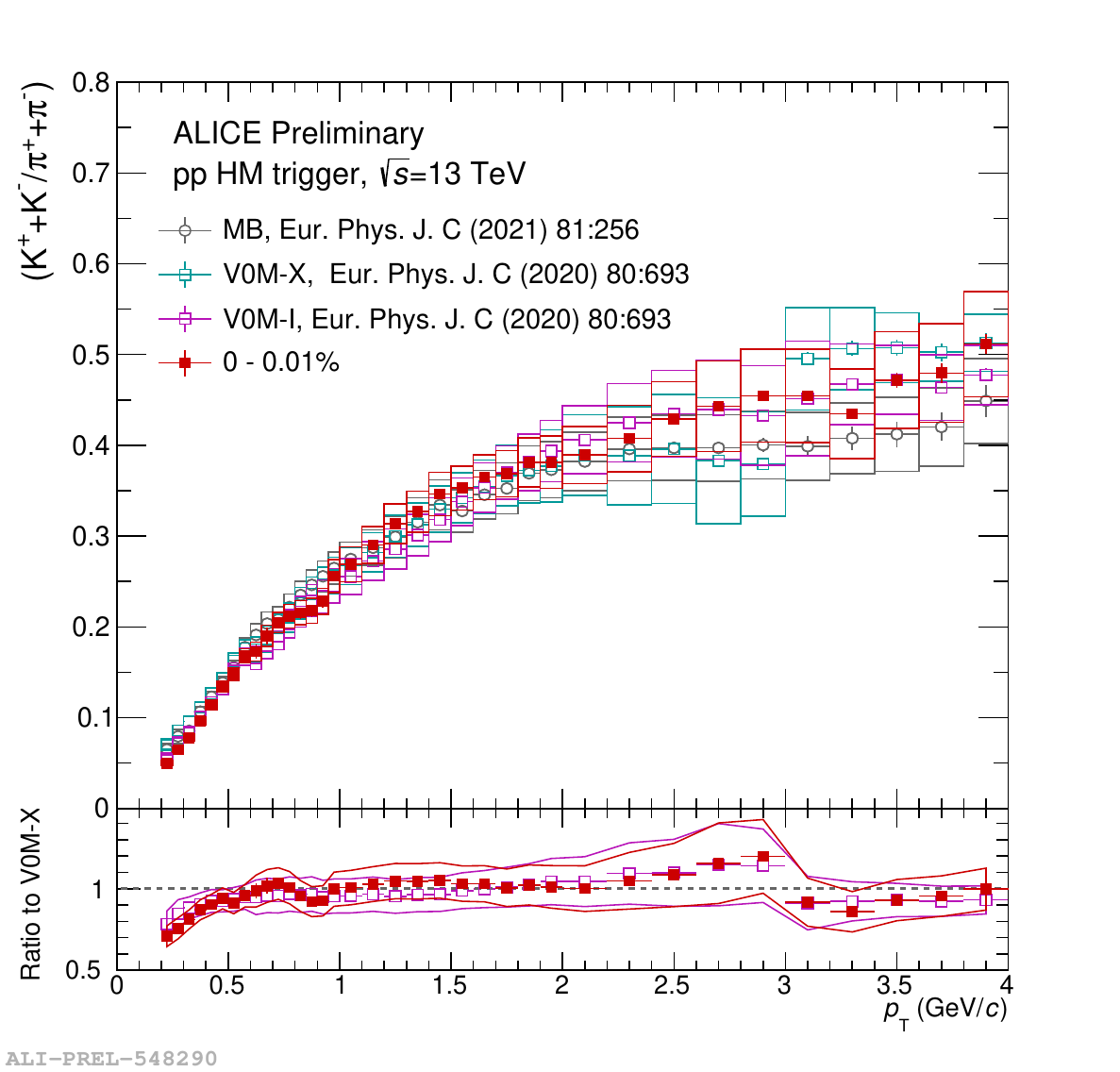}
	\includegraphics[width=0.45\textwidth]{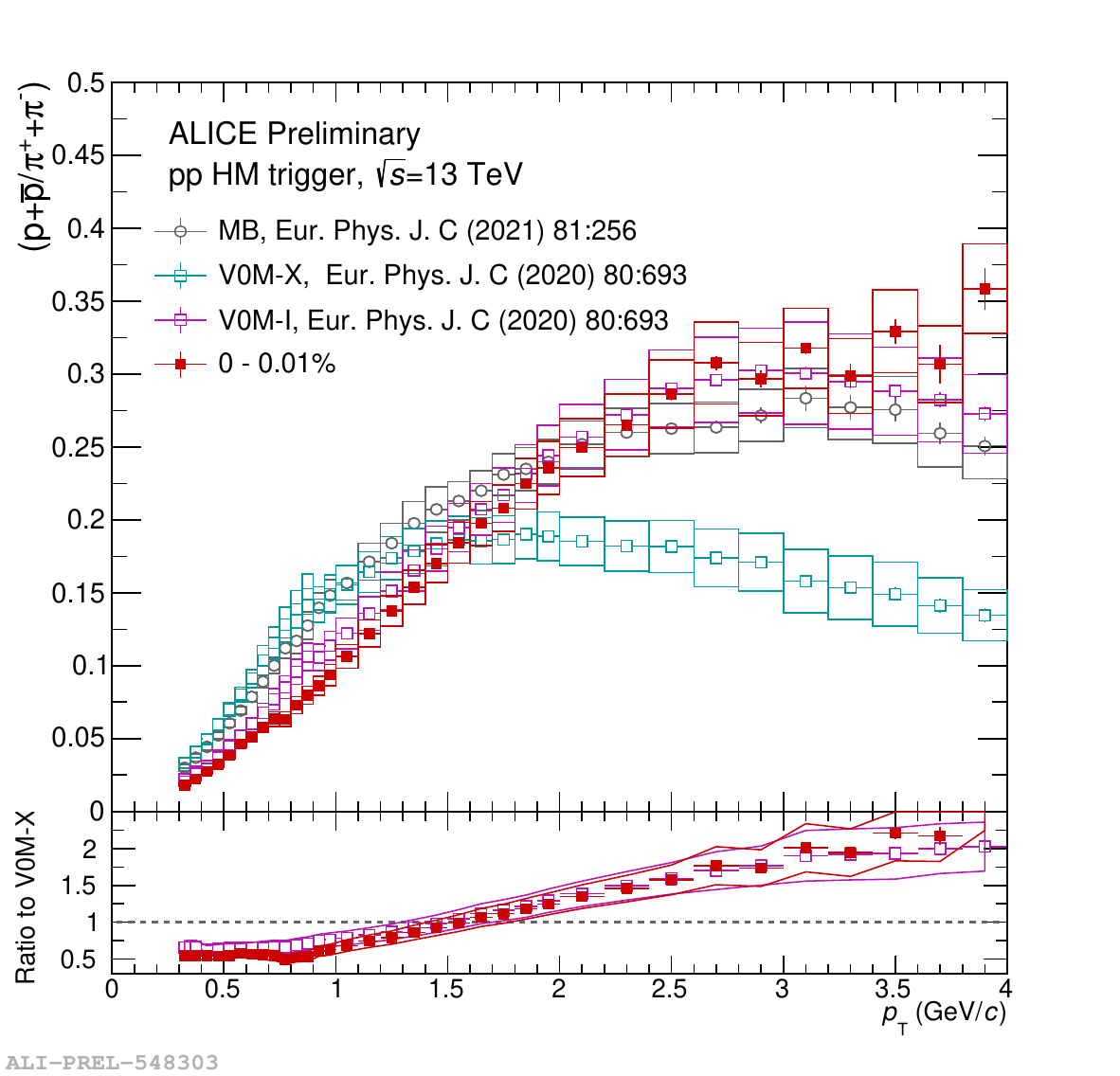}
	
	\caption{Ratios ${\rm K}/{\pi}$ (left) and ${\rm p}/{\pi}$ (right) of the $p_{\rm T}$-spectra. The reference results of the V0M and minimum bias classes for comparison \cite{pp_13TeV_2020, pp_7_13_TeV_MB_2021}. The ratio to V0M-X are shown at bottom panels.}
	\label{fig:ratio_part_pT_HM1}
\end{figure} 

The ratios ${\rm K}/{\pi}$ and ${\rm p}/{\pi}$ as a function of $p_{\rm T}$ are shown in Fig.~\ref{fig:ratio_part_pT_HM1} for the highest multiplicity class (0-0.01\%). The results are compared with the V0M and minimum bias (MB) multiplicity classes for the same collision system (pp) and centre of mass energy ($\sqrt{s}$ = 13 TeV) \cite{pp_13TeV_2020, pp_7_13_TeV_MB_2021}. The lower panels of the figure show a ratio comparison to the V0M-X class. The ${\rm K}/{\pi}$ ratio of this analysis follows the results of the V0M-I class and there is no significant change within uncertainty when comparing to V0M-X over the entire $p_{\rm T}$ range. The ${\rm p}/{\pi}$ ratio exhibits a similar enhancement as seen for V0M-I class with respect to the V0M-X class with increasing $p_{\rm T}$. Such an increase can be explained as due to radial flow boosting the particles with larger rest mass to higher $p_{\rm T}$. A similar result for the ratios ${\rm K}/{\pi}$ and ${\rm p}/{\pi}$ for the other high-multiplicity classes 0.01-0.05\% and 0.05-0.1\% is also found.

\begin{figure}[h]
	\centering
	\includegraphics[width=0.6\textwidth]{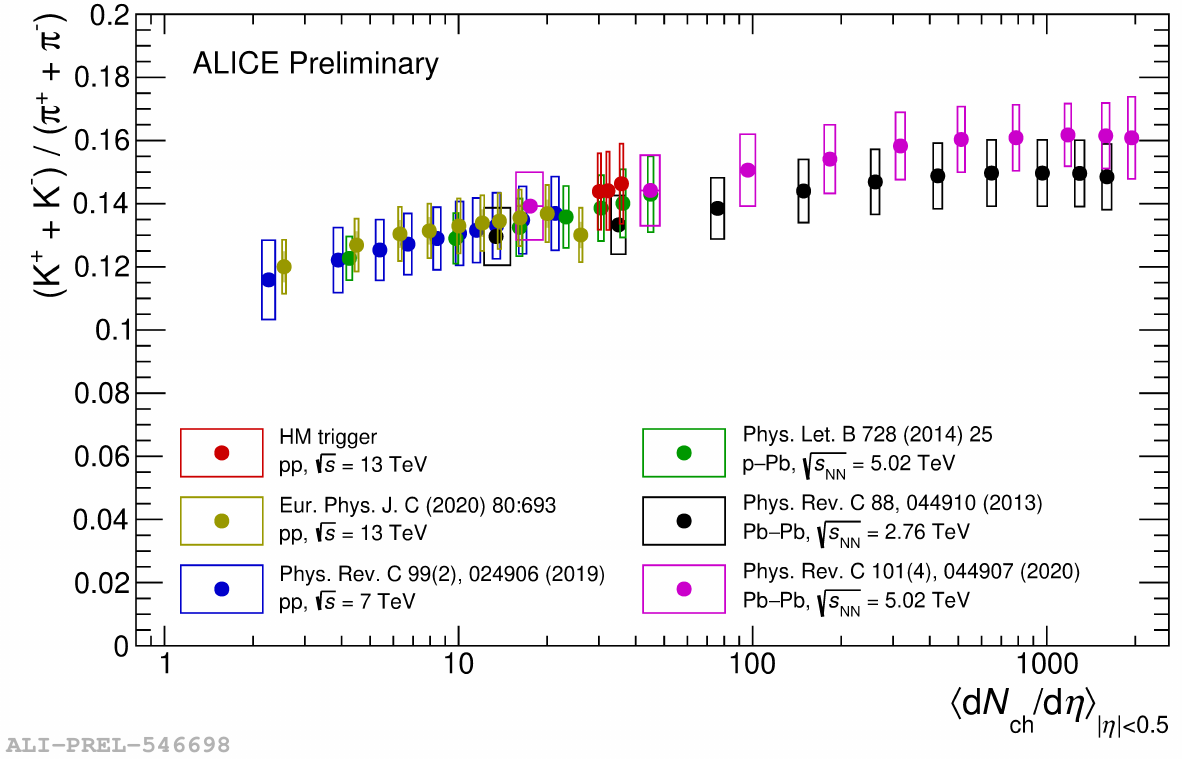}
	\includegraphics[width=0.6\textwidth]{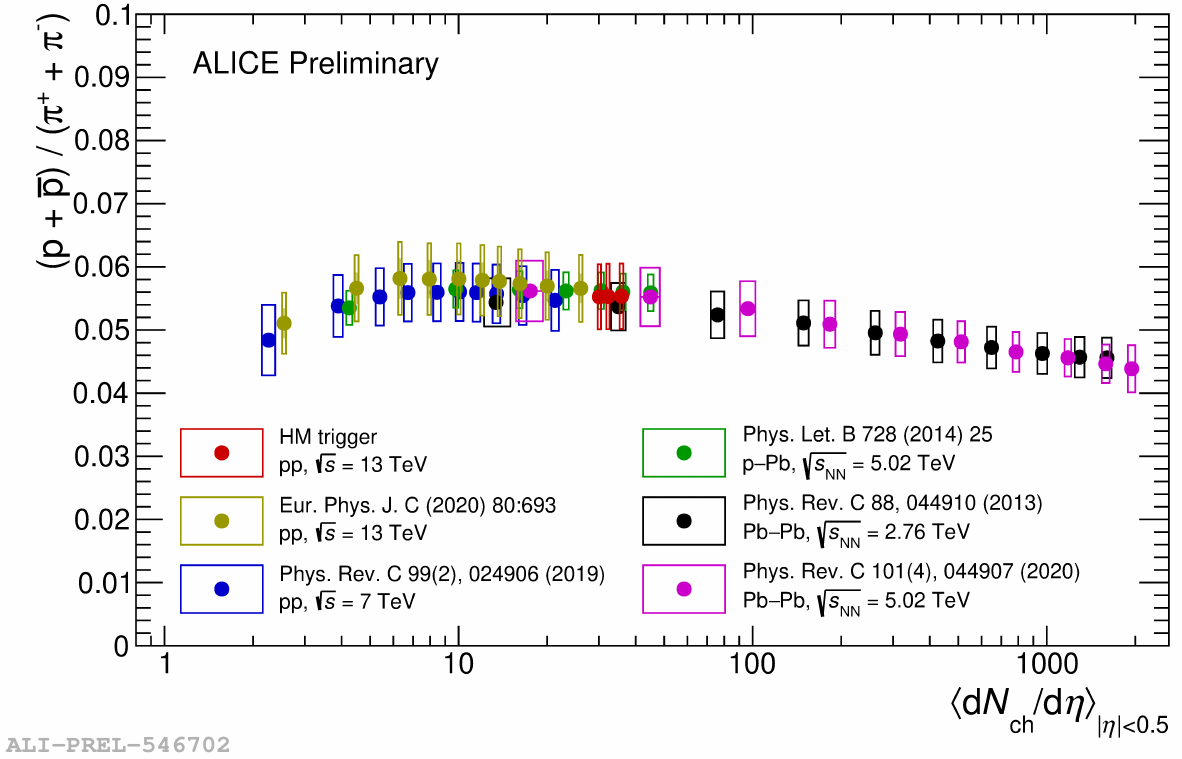}	
	\caption{Integrated yield ratios ${\rm K}/{\pi}$ and ${\rm p}/{\pi}$ as a function of charged-particle multiplicity density ($\langle {\rm d} N_{\rm ch}/{\rm d} \eta \rangle$) of high-multiplicity pp collisions at $\sqrt{s}$ = 13 TeV along with pp, p--Pb and Pb--Pb results \cite{pp_13TeV_2020, pp_7_TeV_2019, pPb_5020GeV_2014, PbPb_2760GeV_2013,  PbPb_5020GeV_2020}.}
	\label{fig:integrated_particle_ratio}
\end{figure}

The ratios of $p_{\rm T}$-integrated yields of kaon- and proton-to-pion are measured in high-multiplicity pp collisions at $\sqrt{s}$ = 13 TeV as shown in Fig.~\ref{fig:integrated_particle_ratio} along with the published pp, p--Pb and Pb--Pb results \cite{pp_13TeV_2020, pp_7_TeV_2019, pPb_5020GeV_2014, PbPb_2760GeV_2013,  PbPb_5020GeV_2020}. The high-multiplicity data points are along the trend of the p--Pb and Pb--Pb points. The ratios neither depends on the collision energy nor the colliding system. It shows a smooth evolution with the multiplicity, indicating that multiplicity is the driving force of hadrochemistry. A small increasing of ${\rm K}/{\pi}$ ratio with increasing multiplicity supports the picture of an increasing strange quark production with increasing multiplicity. A decreasing trend of the ${\rm p}/{\pi}$ ratio with multiplicity is consistent with the hypothesis of antibaryon-baryon annihilation in the hadronic phase.
\section{Conclusion}
The transverse momentum spectra of the $\pi$, K and p are studied in high-multiplicity pp collisions at $\sqrt{s}$ = 13 TeV. The multiplicity and mass-dependent hardening of the $p_{\rm T}$-spectra is observed as found in p--Pb and Pb--Pb system that is well explained by the radial flow effect \cite{pPb_5020GeV_2014, PbPb_2760GeV_2013, PbPb_5020GeV_2020}.  An increasing trend of the mean transverse momentum $\langle p_{\rm T} \rangle$ with multiplicity can be explained by a higher flow at larger multiplicity. The mass ordering of steepness of $\langle p_{\rm T} \rangle$ is understood as larger flow effect on massive particles. The ratios ${\rm K}/{\pi}$ and ${\rm p}/{\pi}$ of the $p_{\rm T}$-spectra confirm similar trend as found in the V0M-I multiplicity class of pp collisions at $\sqrt{s}$ = 13 TeV \cite{pp_13TeV_2020}. The ratios of $p_{\rm T}$-integrated yields of ${\rm K}/{\pi}$ and ${\rm p}/{\pi}$ in Fig.~\ref{fig:integrated_particle_ratio} show a smooth variation with multiplicity for all pp, p--Pb and Pb--Pb systems. The comparison of data across different collision systems, therefore, exhibit collectiveness can be commonly found in both small and large systems.

\end{document}